\documentstyle[emulateapj,psfig]{article}

\begin{document}

\submitted{ApJ accepted May 8 2000}

\title{High-Resolution Imaging of Molecular Gas and Dust in the 
Antennae (NGC 4038/39): Super Giant Molecular Complexes}

\author{Christine D. Wilson\altaffilmark{1,2},
Nicholas Scoville\altaffilmark{2},
Suzanne C. Madden\altaffilmark{3},
Vassilis Charmandaris\altaffilmark{4,5}}

\altaffiltext{1}{Department of Physics and Astronomy, McMaster University,
Hamilton, Ontario L8S 4M1 Canada} 
\altaffiltext{2}{Division of Physics, Mathematics, and Astronomy,
Caltech 105-24, Pasadena CA 91125 U.S.A.}
\altaffiltext{3}{CEA/DSM/DAPNIA/Service d'Astrophysique, 
CE-Saclay, 91191 Gif sur Yvette Cedex, France}
\altaffiltext{4}{DEMIRM, Observatoire de Paris, 61 
Avenue de l'Observatoire, F-75014 Paris, France}
\altaffiltext{5}{Astronomy Department, Cornell University, Ithaca NY 14853 
U.S.A.}

\begin{abstract}

We present new aperture synthesis CO maps of the Antennae (NGC 4038/39) 
obtained with the Caltech Millimeter Array. These sensitive images
show molecular emission associated with the two nuclei and a partial
ring of star formation to the west of NGC 4038, as well as revealing
the large extent of the extra-nuclear region of star formation
(the ``overlap region''), which
dominates the CO emission from this system. The largest molecular
complexes have masses of $3-6 \times 10^8$ M$_\odot$, 
typically an order of magnitude larger than the largest structures seen
to date in more quiescent galaxy disks. The extremely red luminous
star clusters identified previously with HST are well-correlated with
the CO emission, which supports the conclusion that they are highly
embedded young objects rather than old globular clusters. 
There is an
excellent correlation between the CO emission and the 15 $\mu$m
emission seen with ISO, particularly for the brightest regions.
The most massive complexes
in the overlap region have similar [NeIII]/[NeII]
ratios, which implies that all these regions are forming many massive stars.
However, only 
the brightest mid-infrared peak shows strong, rising {\it continuum}
emission longward of 10 $\mu$m, indicative of 
very small dust grains heated to high temperatures by their
proximity to nearby luminous stars. 
Since these grains are expected to be removed
rapidly from the immediate environment of the massive stars, it is
possible that this region contains very young ($< 1$ Myr)
sites of star formation. Alternatively,  fresh dust
grains could be driven into the sphere of influence of the massive
stars, perhaps by the bulk motions of two giant molecular complexes.
The kinematics and morphology of the CO emission in this region provide
some support for this second scenario.

\end{abstract}

\keywords{galaxies: individual (NGC 4038, NGC 4039) -- galaxies: ISM --
ISM: molecules}

\section{Introduction}

The all-sky survey carried out by the Infrared Astronomical Satellite 
identified a large number of galaxies that are extremely luminous in
the infrared (i.e. \markcite{s84}Soifer et al. 1984). 
Follow-up optical imaging revealed that galaxy interactions
play a key role in generating these large infrared luminosities, since 
virtually
all of the nearby galaxies with infrared luminosities $> 10^{12}$ L$_\odot$
are strongly interacting or merger systems 
(\markcite{s88}Sanders et al. 1988). Recent spectroscopic work
with the Infrared Space Observatory (ISO) suggests that a major fraction of the
infrared luminosity in these systems is generated by a starburst
rather than an active galactic nucleus (\markcite{g98}Genzel et al. 1998;
\markcite{l98}Lutz et al. 1998; \markcite{l99}Laurent et al. 2000).
Identifying the origin of the intense star formation produced in galaxy
mergers is also  an important step towards understanding galaxy formation
in the early universe, since the faint submillimeter
sources seen at high redshift 
are similar in many ways to the ultraluminous
infrared galaxies seen in the local universe (\markcite{i98}Ivison et al. 
1998; \markcite{b99}Blain et al. 1999).

At a distance of only 19 Mpc ($H_o = 75$ km s$^{-1}$ Mpc$^{-1}$), 
the Antennae system (NGC 4038/39, Arp 244) is
the closest example of a major merger between two gas-rich spiral galaxies.
Thus, the Antennae provide us
with a unique opportunity to study the
internal structure of a merger in progress.
It has an infrared luminosity of $10^{11}$ L$_\odot$, sufficient
for it to be classified as a luminous infrared galaxy (\markcite{sm96}Sanders
\& Mirabel 1996). 
Early CO observations identified three
large concentrations of molecular gas, one in each of the two nuclei and
a third, more massive concentration in an intermediate region dubbed the 
``overlap region'' (\markcite{s90}Stanford et al. 1990), while the 
total molecular
gas content of this system has been estimated to be $\sim 10^{10}$ M$_\odot$
(\markcite{g98}Gao et al. 1998).
Recent 850 $\mu$m continuum data also show emission from the two nuclei
and the overlap region (\markcite{h00}Haas et al. 2000).
Observations with the Hubble Space Telescope have
identified a population of hundreds of compact, luminous star clusters,
which are likely the young counterparts of present-day globular clusters
(\markcite{ws95}Whitmore \& Schweizer 1995; \markcite{w99}Whitmore
et al. 1999). These star clusters have masses in the range of $10^4$ to $10^6$
M$_\odot$ and follow a power-law mass spectrum (\markcite{z99}Zhang
\& Fall 1999).
Recent observations with ISO in the mid-infrared 
have identified an extremely
luminous region of massive star formation that is almost completely obscured
at optical wavelengths (\markcite{v96}Vigroux et al. 1996). This compact region
is so luminous that it accounts for 15\%  of the 15 $\mu$m
flux of the entire system (\markcite{m98}Mirabel et al. 1998). In fact,
the Antennae is the only nearby interacting system in which the most
luminous mid-infrared region is not one of the two nuclei
(\markcite{x00}Xu et al. 2000).

Since stars form from molecular gas, determining the physical properties
and kinematics of the molecular gas in merger systems is important
for understanding how star formation is triggered
by galaxy mergers. In this paper, we present new, more sensitive,
high-resolution CO observations of the Antennae made with
the Caltech Millimeter Array. These
observations cover most of the inner regions of the disks, which
contain the luminous star clusters and bright mid-infrared sources.
We compare these CO data to HST images from \markcite{w99}Whitmore
et al. (1999) and to mid-infrared imaging and spectroscopic data from 
ISO (\markcite{v96}Vigroux et al. 1996; \markcite{m98}Mirabel et al. 1998).
The observations and data reduction are presented in \S 2, while the
properties of the most massive molecular complexes are discussed
in \S3. In \S 4, we focus on the mid-infrared properties of these
massive complexes, and, in particular, on the origin of the steeply
rising continuum emission, which is a unique feature of the most luminous
region.
The paper is summarized in \S 5. A more detailed discussion
of the entire CO data set, including a comparison with the young star clusters 
identified with HST and an analysis of possible scenarios for star
cluster formation in light of the new CO data, 
will be presented in a second paper (Paper II, Wilson et al., in preparation).

\section{Observations and Data Reduction}

Observations of three overlapping fields in the Antennae were obtained with
the Caltech Millimeter Array between 1998 March and 1999 February.
The primary beam of the array at this
frequency is 66$^{\prime\prime}$ and the field centers were separated by
40$^{\prime\prime}$. The phase calibrator for these observations was
3C279 and the flux calibrator was 3C273, which was itself calibrated against
Uranus and Neptune using the standard Caltech Millimeter Array database.
Observations of each field were obtained in four configurations of the array,
which resulted in a synthesized beam of $3.15 \times 4.91 ^{\prime\prime}$
or $310 \times 480$ pc at the distance of the Antennae.
Due to the low declination of the source, the typical 
single sideband system temperatures ranged between 1000 and 2000 K, which
gave an rms noise in the maps of 0.055 Jy beam$^{-1}$ (0.33 K) in a
5.2 km s$^{-1}$ (2 MHz) channel. 

The data were mapped and analyzed using the MIRIAD package 
(\markcite{s95}Sault, Teuben, \& Wright 1995). Data for the three fields
were combined with robust weighting 
to produce a single mosaic data cube and the data were
cleaned using the task MOSSDI with a cutoff of
2$\sigma$ (0.11 Jy beam$^{-1}$). Individual molecular complexes were 
identified from the data cube using the automatic clump identification
algorithm CLFIND (\markcite{w94}Williams, de Geus, \& Blitz 1994).
This algorithm searches for peaks of emission within a contour map
of the data, which it then follows down to lower intensity levels,
and has the advantage of not assuming any specific clump profile
(such as a gaussian profile). We used 
a contour level of 2$\sigma$ (0.11 Jy beam $^{-1}$). The individual
complexes were inspected by eye and, in several cases, two or more
complexes that appeared to belong to a single structure rather than to
be separate objects were merged by hand. The properties of
the individual complexes were measured using the program CLSTATS
(\markcite{w94}Williams, de Geus, \& Blitz 1994). This program
combines the data cube with the clump assignment cube produced
by CLFIND to calculate individual statistics
(such as total flux, velocity width, and radius). The
data were corrected for the varying primary beam attenuation
across the field by dividing by the gain image; regions of
the mosaic map with a primary beam attenuation
lower than 0.5 were masked. An integrated intensity
map was created by summing the emission from all the complexes with
velocity extents of at least 3 channels (15.6 km s$^{-1}$).
More details of the 
identification of the complexes 
and the data reduction will be given in Paper II.

\section{Molecular Gas in the Antennae and the Properties of the Largest 
Complexes}

\begin{figure}[ht] 
\figurenum{1} 
\plotone{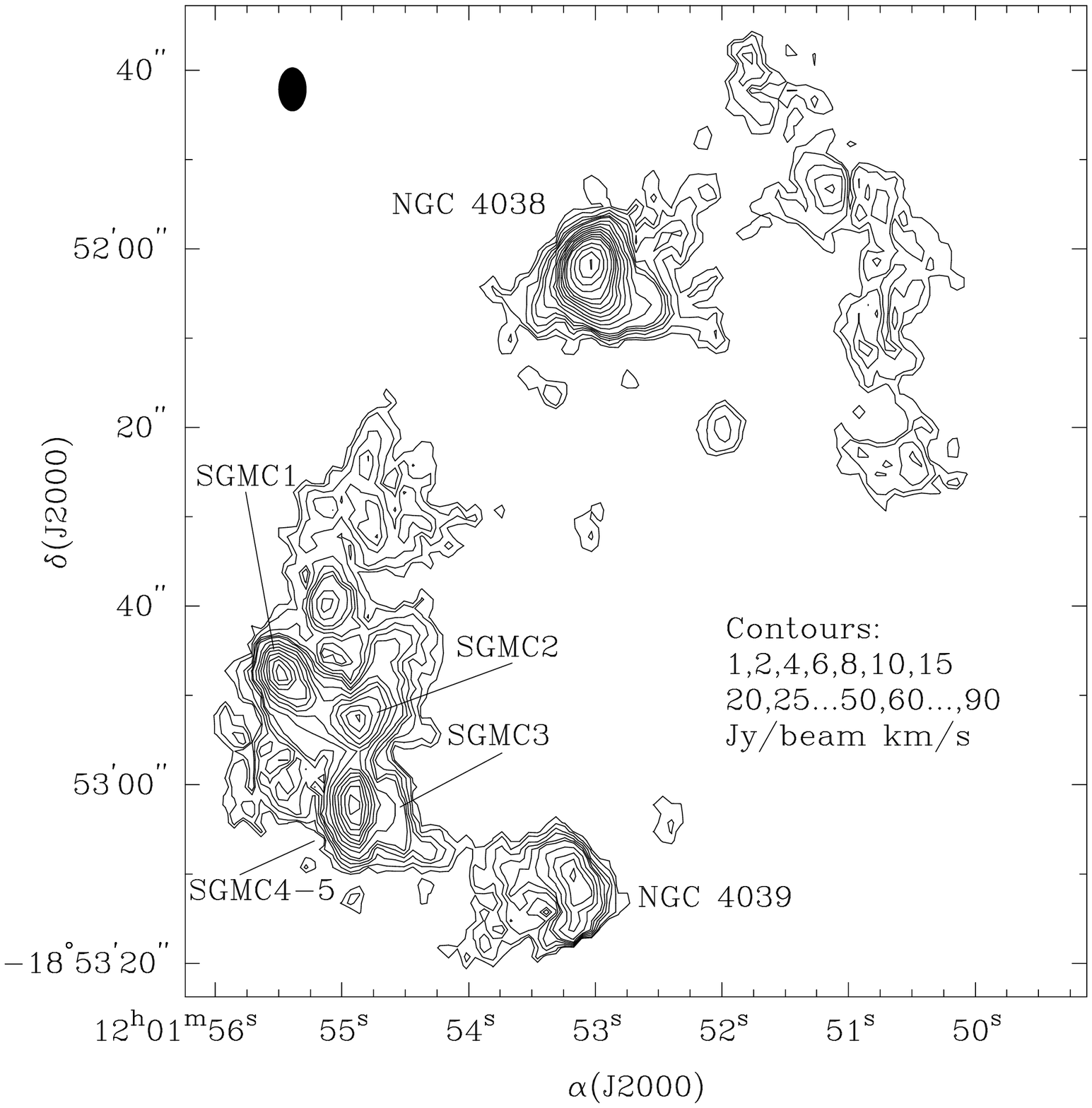}
\caption{The CO integrated intensity map of the 
inner regions of the Antennae. This image is a mosaic of data from
three pointing centers and has been corrected for the sensitivity
fall-off of the primary beam. A $40 \times 50^{\prime\prime}$ region
in the south-west corner of the figure was not imaged in CO. The two
galaxy nuclei and five super giant molecular complexes (SGMCs) in the
``overlap region'' are indicated on the map. The synthesized beam of
$3.15\times 4.91^{\prime\prime}$ is indicated by the filled circle in
the upper left corner.
\label{fig-1}}
\end{figure}

\begin{figure}[ht] 
\figurenum{2} 
\plotone{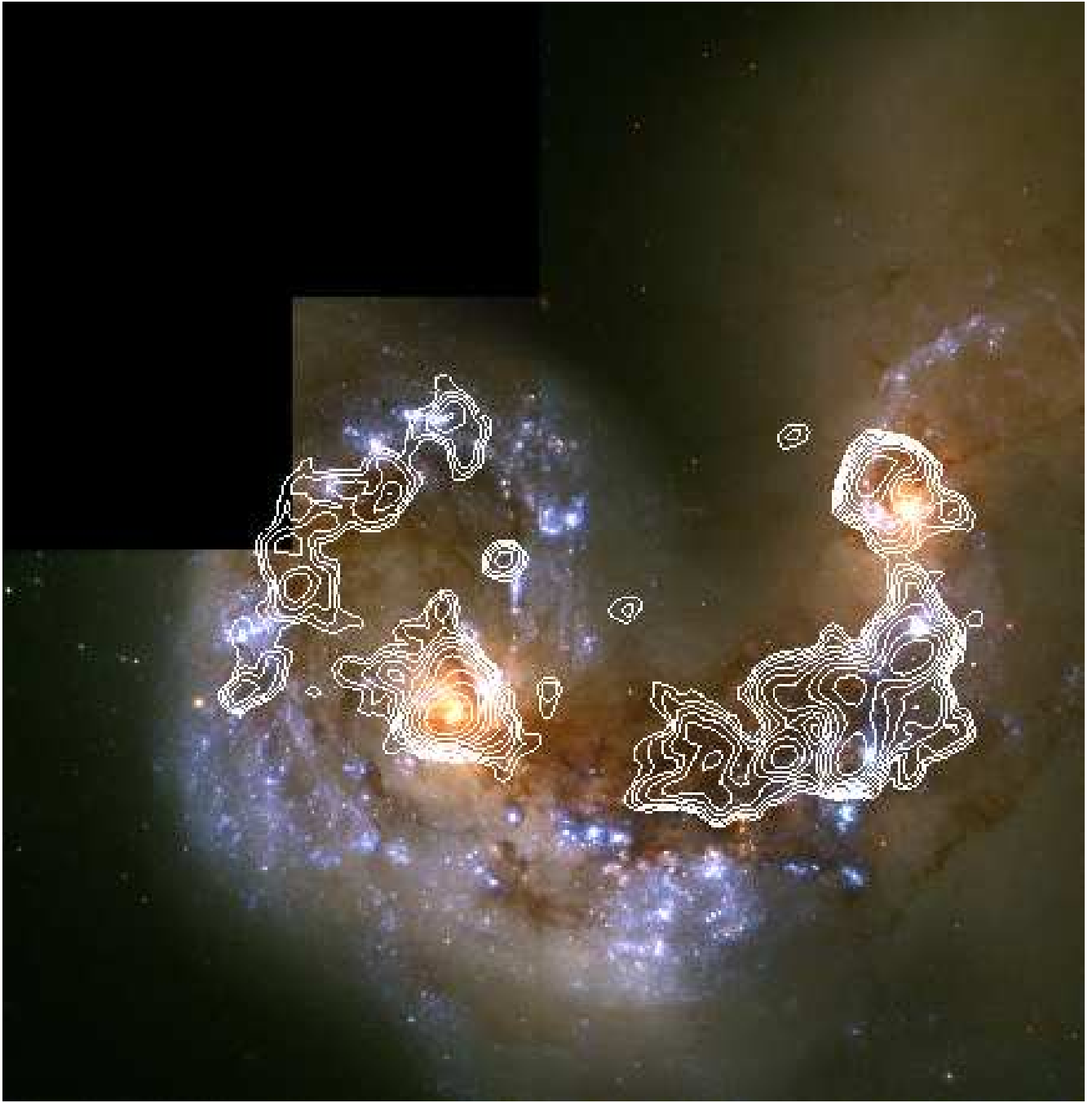}
\caption{The CO integrated intensity map of Figure 1 
overlaid on the true-color HST image from Whitmore et al. (1999).  The
CO image has been convolved with a 2$^{\prime\prime}$ gaussian to
smooth the image slightly.  The contour levels are
1,1.6,2.5,4,6,9.5,15,23,37,57 percent of the peak intensity.
\label{fig-2}}
\end{figure}

The CO integrated intensity map is shown in Figure~\ref{fig-1} and
is overlaid on the HST image
from \markcite{w99}Whitmore et al. (1999) in Figure~\ref{fig-2}. The strongest
CO peak is associated with the nucleus of NGC 4038 in the north. 
Significant CO emission is also associated with the nucleus of NGC 4039,
but fully half the CO emission comes from  a $\sim 3 \times 5$ kpc region
between the two nuclei. The strong emission in the southern half of this
region corresponds to the ``overlap region'' identified by 
\markcite{s90}Stanford et al. (1990). Comparison with the map of
\markcite{s90}Stanford et al. (1990) shows that our new map detects
significantly more extended CO emission than the previous map; indeed,
the entire northern half of the extended overlap region is not visible
in the older maps. Because of our larger area coverage, we also detect
significant CO emission associated with the blue arc of star formation
located to the west of NGC 4038. 

The total flux detected in our map
is 910 Jy km s$^{-1}$, compared to a total flux of 430 Jy km s$^{-1}$
in the map of \markcite{s90}Stanford et al. (1990). 
Most of the increase
in detected flux is due to the higher sensitivity of the new map (due to
the larger number of telescopes and more sensitive receivers). In addition, a
more compact configuration was included in these observations, which gives 
better sensitivity to structures on larger scales. 
Gao et al. (1998) have made an extensive single dish map of the
Antennae, which detects five times more flux 
than \markcite{s90}Stanford et al. (1990).
If all this flux were to lie within
the area we have mapped, then we would have detected 
40\% of the single dish flux across our entire map. 
In a single 55$^{\prime\prime}$ beam pointed (quite by chance) at 
the overlap region, \markcite{sm85}Sanders \& Mirabel
(1985) measured an integrated CO intensity of 15.9 K km s$^{-1}$. 
Adopting a 
CO-to-H$_2$ conversion factor of
$3 \times 10^{20}$ H$_2$ cm$^{-2}$ (K km s$^{-1}$)$^{-1}$ 
(\markcite{strong88}Strong et al. 1988) and 
including a factor of 1.36 to account for heavy elements, this intensity
corresponds to  $2.6 \times 10^9$ M$_\odot$ of
molecular gas. In the same region of our map, we detect 470 Jy km s$^{-1}$
for a total mass of $2.7 \times 10^9$ M$_\odot$. Thus, at least in the
overlap region, we likely detect most of the molecular gas present.
For the individual complexes discussed below,
nearly all of the flux should be present in our map since they are much 
smaller than the $\sim 20^{\prime\prime}$ spatial scale at which our 
interferometer data resolves out structures. The residual single dish 
flux must be more extended or at lower surface brightness levels than
the emission in our map. The total flux detected in our map corresponds
to a mass of $5.3\times 10^9$ M$_\odot$ of molecular gas.

The physical properties of the seven largest molecular complexes seen in our
map are given in Table~\ref{tbl-1} and their locations are identified
in Figure~\ref{fig-1}. The molecular and virial masses were
calculated from the CO flux, diameter, and velocity width using the equations
given in \markcite{ws90}Wilson \& Scoville (1990). (One difference is
that the diameter is measured from $D =2\sqrt{A/\pi}$, where $A$ is the
area of the cloud at the 2$\sigma$ contour level; the diameter was then
deconvolved from the synthesized beam.) These seven complexes are
significantly larger and more massive than typical giant molecular clouds
seen in the Milky Way, which have masses in the range of $10^4$ to
$10^6$ M$_\odot$ (i.e.
\markcite{sss}Sanders, Scoville, \& Solomon 1985). 
This result is not too surprising, since our spatial resolution and 
sensitivity are insufficient to detect even the most massive Milky Way
cloud at the distance of the Antennae.
However, the complexes are also a
factor of 5-10 times more massive than the most massive giant molecular
associations seen in the grand-design spiral galaxy M51 
($6\times 10^7$ M$_\odot$, \markcite{rk90}Rand \& Kulkarni 1990),
which has been observed with similar resolution and mass sensitivity. 
Because of their unusually large masses, we will refer to these
objects as super-giant molecular complexes, or SGMCs. Note that SGMC4 and
SGMC5 lie at the same location but at different velocities, and so are
not visible as separate emission peaks in Figure~\ref{fig-1}.
Cross-identification with previous work is given in Table~\ref{tbl-2}.
Most of the complexes have molecular masses which
are comparable to or larger than their virial masses, which suggests that
these complexes could be gravitationally bound. Only the nucleus of
NGC 4039 has a molecular mass that is significantly smaller than its virial
mass; however, since this complex is located in a galactic nucleus,
there is likely to be a significant mass contribution by the stars in both
the disk and the bulge, which will affect the kinematics of the molecular
gas. 

Approximately 600 of the 728 star clusters identified by 
\markcite{ws95}Whitmore \& Schweizer (1995), including
9 of the 12 very red clusters, lie within the region mapped  in CO. 
Six of the very red star clusters
lie within the CO contours of Figure~\ref{fig-2}, and four
of these are associated with 
the very strong CO emission of the overlap region. Thus,
the 6\% of the area which contains the strongest CO emission contains 
45\% of the very red clusters. This correlation of the
reddest clusters with the CO emission strongly supports the
argument that the red color of these clusters comes from high 
extinction (\markcite{ws95}Whitmore \& Schweizer  1995).

\section{What is the Origin of the Strong Mid-Infrared Emission in
the Antennae?}

\begin{figure}[ht] 
\figurenum{3} 
\plotone{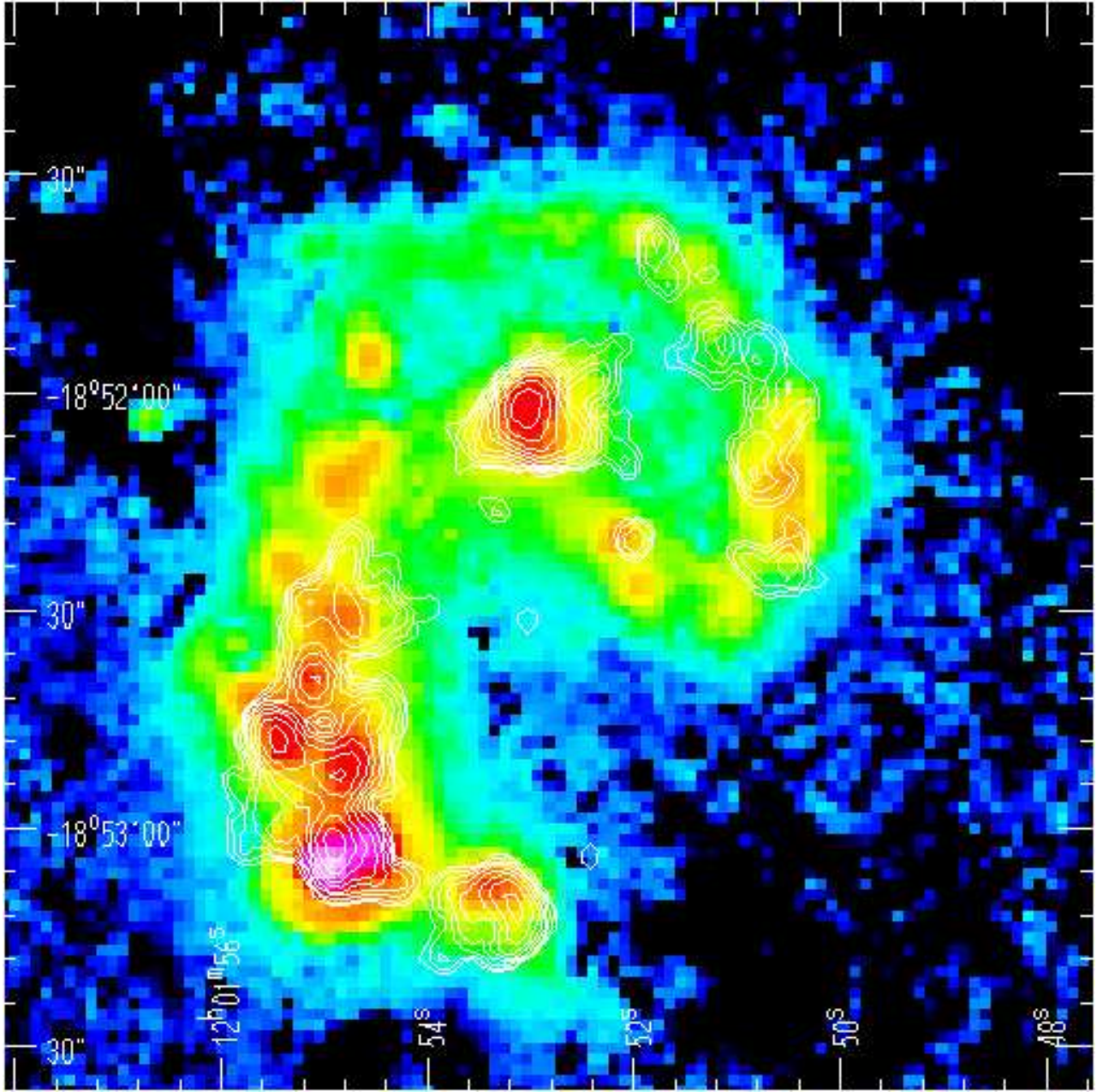}
\caption{The CO integrated intensity map overlaid on the 
ISO 15 $\mu$m image from Mirabel et al. (1998). The CO contours are
1,1.6,2.5,4,6,9.5,15,23,37,57,90 Jy beam$^{-1}$ km s$^{-1}$. Note the
excellent match between the mid-infrared and CO data, particularly
around the strongest CO emission.
\label{fig-3}}
\end{figure}

Comparing our new high-resolution CO map with similarly high-resolution
mid-infrared data from ISO reveals intriguing similarities and
differences among the super-giant molecular complexes in the overlap
region of the Antennae.
Figure~\ref{fig-3} shows the CO contours superimposed on the 15 $\mu$m
(12-18 $\mu$m) ISO image with 4.5$^{\prime\prime}$ resolution 
from \markcite{m98}Mirabel et al. (1998). 
As noted by \markcite{v96}Vigroux et al. (1996), 
the strongest mid-infrared peak is
not associated with either of the two nuclei, but instead lies in the
overlap region. Figure~\ref{fig-3} shows that this strong mid-infrared
peak is associated with two super-giant molecular complexes (SGMC4 and SGMC5).
The other two mid-infrared peaks in the overlap region are associated
with SGMC1 and SGMC2, respectively. In general, there is an excellent
correspondence between the strongest CO peaks and the strongest mid-infrared
peaks.  

\begin{figure}[ht] 
\figurenum{4} 
\plotone{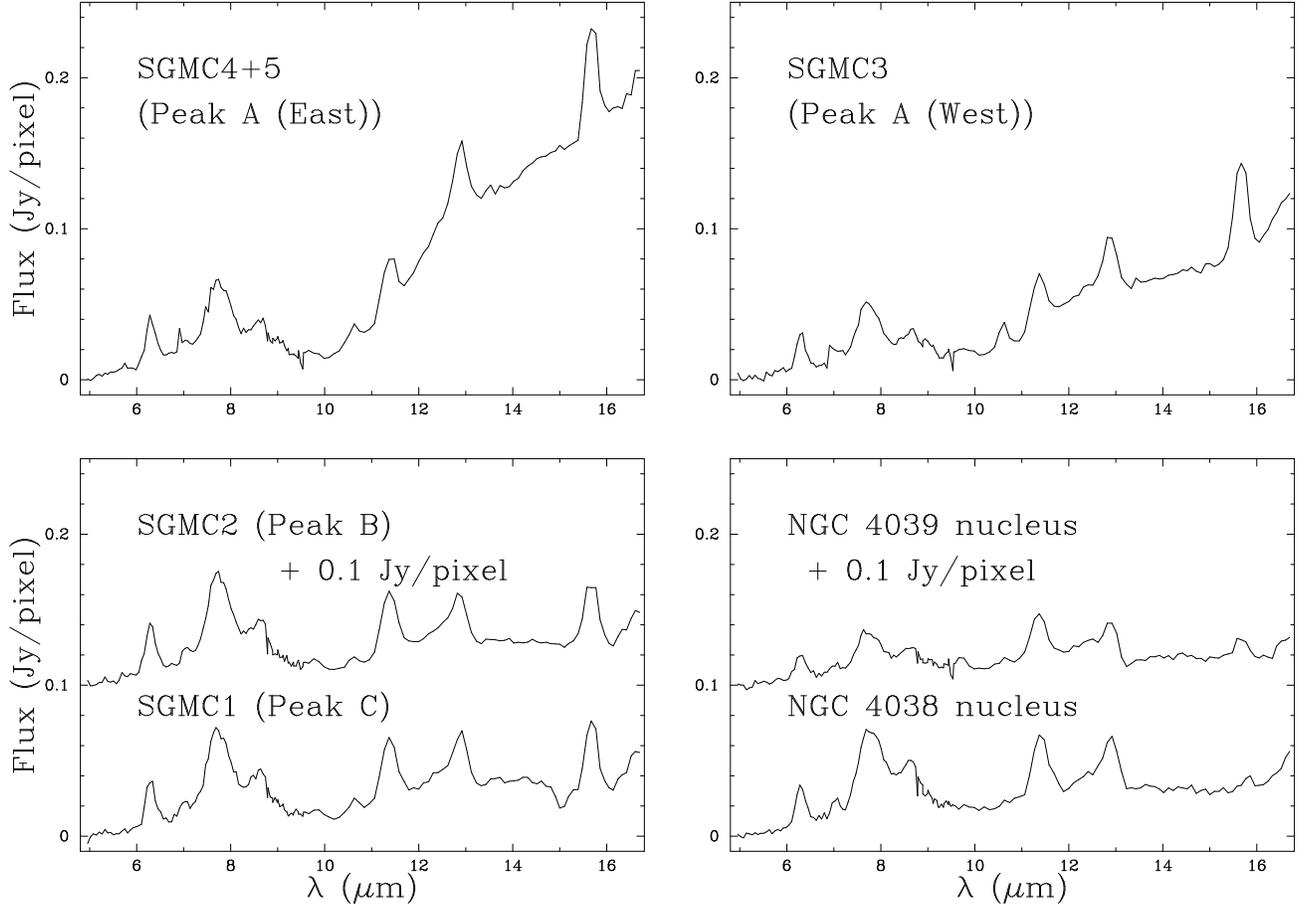}
\caption{ISOCAM CVF spectra for the four main mid-infrared
peaks in the overlap region. Each spectrum is measured from a single
6$^{\prime\prime}$ pixel of the CVF data cube.  The spectra are
identified by their associated molecular gas structure, and also by
the original identification from Vigroux et al. (1996).  The spectra
of the two nuclei are given for comparison; spectra for the nuclei and
the strongest 15 $\mu$m peak (Peak A (East)) have been published
previously by Vigroux et al. (1996; see also Vigroux 1999) and Mirabel
et al. (1998). Note the strong [NeIII] lines at 15.6 $\mu$m seen in
all spectra except those of the two nuclei; the strength of this line
indicates a significant presence of massive stars. Also note the
strong rising continuum emission seen at the longer wavelengths in the
spectra from the SGMC4-5 region.
\label{fig-4}}
\end{figure}

Differences between the complexes begin to be apparent when
we examine their mid-infrared spectra.
Figure~\ref{fig-4} shows the mid-infrared
spectra from 5 $\mu$m to 17 $\mu$m at the positions of each of the
seven SGMCs. While all the SGMCs in the overlap region show strong
emission lines of both [NeIII] (15.5 $\mu$m) and [NeII]
(12.8 $\mu$m), the mid-infrared spectra
of the two nuclei have at most weak [NeIII] emission (see
also \markcite{v96}Vigroux et al. 1996; \markcite{m98}Mirabel
et al. 1998; \markcite{v99}Vigroux 1999).
The strong [NeIII] lines and large [NeIII]/[NeII] ratios\footnote[1]{We 
note that, given the spectral resolution of
the ISOCAM CVF, the [NeII] line is blended with the weak
12.7 $\mu$m Unidentified Infrared Band (UIB).
However, since we are concerned with the variation of this
ratio over the various peaks in the overlap region, we can make the
assumption that the relative strengths of the UIB features are
approximately invariant in the interstellar medium (see also 
\markcite{b99}Boulanger 1999; \markcite{u00}Uchida et al. 2000). 
Thus, when we state
the [NeIII]/[NeII] ratio  (Table~\ref{tbl-2}), 
we are always referring to the blended [NeII] line.
We expect the contamination by the 12.7 $\mu$m UIB to result in
an underestimate of the true [NeIII]/[NeII] ratio by no more than 20\%
(i.e. \markcite{u00}Uchida et al. 2000).}
throughout the overlap region (Table~\ref{tbl-2})
indicate that all five molecular complexes in the overlap region contain
embedded regions of massive star formation.
The observed line ratios correspond to effective temperatures 
of $4.4-4.5 \times 10^4$ K (\markcite{k96}Kunze et al.
1996) or O stars with masses of 40-60 M$_\odot$ and main
sequence lifetimes of $\sim 4$ Myr (\markcite{s96}Schaerer 
et al. 1996). This high effective temperature has also been confirmed by
observations of the He I/Br$\gamma$ ratio for
the bright cluster in the SGMC4-5 region (\markcite{g00}Gilbert et al. 2000).

Although the relative strengths of the neon lines and the UIBs in
the various molecular complexes in the overlap region are quite
similar to one another, there is a striking difference in the mid-infrared
{\it continuum} emission between the different complexes. In particular,
SGMC4-5 (and to some extent SGMC3) show strong continuum emission
that rises towards longer wavelengths, while the continuum emission in SGMC1
and SGMC2 is essentially flat (Figure~\ref{fig-4}). 
The net effect of this rising spectrum is to increase the flux of
SGMC4-5 seen in the broad-band 15 $\mu$m filter (Figure~\ref{fig-3}),
which gives it enhanced 15$\mu$m/7$\mu$m and 15$\mu$m/CO ratios compared
to the other two regions (Table~\ref{tbl-2}). Thus, most of
the high luminosity of this region seen in the 15 $\mu$m image is due to
the presence of this unusual continuum spectrum. A rising
mid-infrared continuum spectrum, such as is seen in SGMC4-5, can be 
produced by a population of very small dust grains that are heated
to quite high temperatures (of the order of 100 K) due to their proximity
to massive O stars. In the Milky Way, this rising continuum emission 
is seen to be closely 
confined to HII regions and photon-dominated regions; for example,
in M17, the strong continuum emission is confined to a region roughly
1 pc in extent (\markcite{ver96}Verstraete et al. 1996; \markcite{c96}Cesarsky
et al. 1996). Thus, to have strong mid-infrared continuum emission
requires the presence of very small dust grains within a few parsecs
of the massive stars.

These observations of the overlap region of the Antennae present us with
an interesting puzzle. The three bright mid-infrared peaks seen in
Figure~\ref{fig-3}, which correspond to SGMC1, SGMC2, and SGMC4-5, contain
similar masses of molecular gas, and their neon lines indicate that they 
contain similar populations of massive stars. Yet one region (SGMC4-5)
contains strong and rising mid-infrared continuum emission, which indicates
the presents of very warm and small dust grains, while the other
two regions have much flatter mid-infrared spectra (Figure~\ref{fig-4}).
Why does the SGMC4-5 region contain a population of warm dust grains which
the other two, otherwise similar, regions lack?
O stars are highly effective at modifying their immediate environment
by means of their high luminosities and strong stellar winds. Models
and observations of superbubbles produced by these strong stellar
winds show that a single
massive O star can drive a stellar
wind bubble to a radius of 20 pc in about 1 Myr 
(\markcite{o95}Oey \& Massey 1995). 
Thus, only younger regions would likely have dust grains located 
close enough to the massive stars to produce a rising mid-infrared
continuum spectrum. Alternatively, for an older region 
to produce such a strong mid-infrared continuum, it would need
its supply of dust grains replenished by 
some mechanism, perhaps infall of new material or an encounter with
a second molecular cloud.
Finally, given the relatively large spatial scales ($>$500 pc) of the
regions we are studying, it is also possible that a single SGMC region
contains a number of star formation sites of various ages, and we are
observing dilution effects in our mid-infrared spectra. For example,
in a region with a large number of extremely young sites of massive
star formation, the steep mid-infrared continuum spectrum from these
regions may dominate over the flat spectra of the older
regions to produce a spectrum similar to that seen in the SGMC4-5 region.

The information that is available on the relative ages of the stars
associated with these molecular complexes is inconclusive.
The [NeIII]/[NeII] line ratios indicate similar populations of O
stars in all the regions (Table~\ref{tbl-2}). Since these massive O
stars only live for $\sim 4$ Myr (\markcite{s96}Schaerer et al. 1996),
these observations place an upper limit on the age of the regions.
(However, if these regions are undergoing continuous star formation, they
could be considerably older. In this case, the age would refer to
the most recent episode of star formation.)
The HST data of \markcite{w99}Whitmore et al.
(1999) show very compact
H$\alpha$ bubbles in the vicinity of SGMC1,
which suggests that this complex is relatively young.
Neither SGMC2 nor SGMC4-5 appear to have significant H$\alpha$ emission.
This lack of H$\alpha$ emission could mean that these regions are so
young that the star formation is still deeply cocooned and hence
invisible at optical wavelengths. 
However, all the regions do show Br$\gamma$ 
emission (\markcite{f96}Fischer et al. 1996). 
Using instantaneous starburst models, the intrinsic $V$-band brightness,
and the Br$\gamma$ flux,
\markcite{g00}Gilbert et al. (2000) estimate the age of the brightest
star cluster near SGMC4-5 to be $\sim 4$ Myr.
Similarly, \markcite{m00}Mengel et al. (2000) 
have estimated the ages of the two
star clusters near SGMC4-5 and SGMC3
using the Br$\gamma$ and $^{12}$CO(2-0) equivalent widths.
They estimate ages of 4-6 Myr, and do not find any significant differences
between these two clusters. These age estimates suggest that
we are not viewing SGMC4-5 at a special time in their evolution, when
massive star formation has been going on for less than a million years.
However, given 
that even the near-infrared observations measure the average emission from an
area of $\sim 100$ pc, it is possible that SGMC4-5 contains a mix
of extremely young clusters ($<$1 Myr), which could contribute the
rising mid-infrared continuum spectrum, and somewhat older regions ($\sim
1-4$ Myr), which could dominate the near-infrared emission lines.

\begin{figure}[ht] 
\figurenum{5} 
\plotone{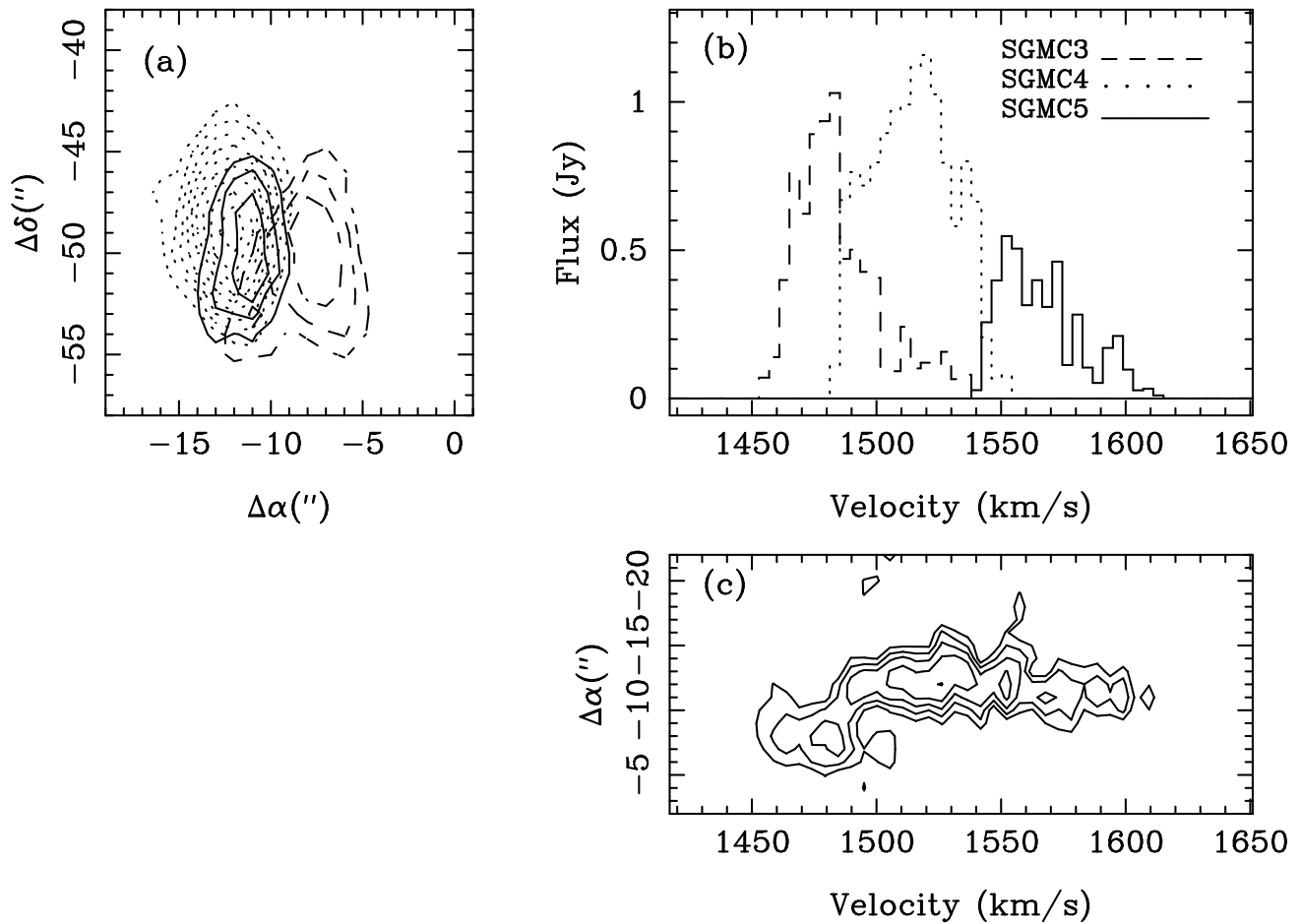}
\caption{(a) Integrated intensity plots for
SGMC3, SGMC4, and SGMC5. Coordinates are offsets in arcseconds from
(12$^h$ 01$^m$ 54.1$^2$, -18$^o$ 52$^\prime$ 13$^{\prime\prime}$). 
Contour levels are the same for all three clouds and are
10,20,30,...,100\% of the peak flux of SGMC4 (39.6 Jy km s$^{-1}$).
SGMC3 is plotted with dashed lines, SGMC4 is plotted with dotted lines, and
SGMC5 is plotted with solid lines. Note that SGMC3 is well offset
spatially from SGMC4. (b) Area-integrated spectra of SGMC3-5. Note 
in particular that SGMC5
is well offset in velocity from SGMC4. (c) Position-velocity plot through
the peak of the emission from SGMC3-5 along an east-west cut. 
Contour levels are 20,40,60,80,100\% of
the peak intensity and the map was smoothed by convolving 
with a 3$^{\prime\prime}$ gaussian. The brightest source is SGMC4. 
The spatial and velocity offset of SGMC3 is
clearly evident. SGMC5 is obvious as a low-emission tail to high velocities
at the same positional offset as SGMC4 (compare with (b) above).
\label{fig-5}}
\end{figure}

The CO data for the SGMC3-5 region suggest that the second scenario,
in which older ($\sim 4$ Myr!) star formation sites are replenished
with dust grains from unprocessed material, deserves a second look. 
One striking aspect of this
region is that it is the only region of strong CO emission (besides the nucleus
of NGC 4039; see Paper II) that is clearly made up of more than a single
molecular complex. In fact, our analysis of this region suggests that
there are three large, gravitationally bound molecular complexes projected
within an area of just 1 kpc$^{2}$ (Table~\ref{tbl-1}). Figure~\ref{fig-5} 
illustrates this in more detail; it is clear that each complex is
well-separated from the other two in velocity but is
close to overlapping spatially. This
close proximity of three massive complexes suggests that cloud collisions 
could be the means of replenishing the small dust grains and
maintaining the rising mid-infrared continuum spectrum in this
region of the Antennae.
In this picture, two or more massive
complexes could collide and form a first generation
of O stars in the shocked boundary layers. The
winds of the O stars would quickly
remove the dust and gas (including the very small grains) from
 their immediate vicinity. However, the
motions of the large amount of remaining material in the two complexes
would continually feed fresh gas and dust into the sphere of influence
of the O stars. Thus, the photon-dominated region around the O stars
would remain out of equilibrium for much longer than would be the
case for a static situation, and hence there would be a significant
population of hot grains to produce the observed mid-infrared continuum
emission over an extended period of time. 
An additional complementary effect can be
the shattering of larger grains in grain-grain collisions in
shocks with velocities of 100 to 150 km s$^{-1}$ to form
the smaller grains that we are seeing in the mid-infrared 
(\markcite{j96}Jones, Tielens, \& Hollenbach  1996).
Such shocks might be expected to produce near-infrared H$_2$ emission lines;
indeed, H$_2$ emission has been seen in the vicinity of SGMC4-5, as well
as within SGMC1 and SGMC2 (\markcite{f96}Fischer et al. 1996). However,
high-resolution near-infrared spectra of the brightest star cluster in 
the SGMC4-5 region reveals almost pure ultraviolet fluorescence
(\markcite{g00}Gilbert et al. 2000).

\section{Conclusions}

We have presented a new, high-resolution CO map of the inner
region of the Antennae (NGC 4038/39) obtained with
the Caltech Millimeter Array. These new data cover a larger
area than previously published aperture synthesis maps (\markcite{s90}Stanford
et al. 1990) and have considerably higher sensitivity and resolution as
well. 

(1) We detect twice as much flux in our new map as was seen in the
original interferometric map by \markcite{s90}Stanford et al. (1990),
and 40\% of the total CO flux seen in the extended single dish map
of \markcite{g99}Gao et al. (1998).
Comparison with older single dish data suggests we may be detecting all of the
flux in the key ``overlap region''. These new data reveal that the overlap
region is even larger than seen previously, 
at least $3\times 5$ kpc in size, and accounts for 50\%
of the molecular gas content of the inner disk of the Antennae. We also detect
CO emission associated with a partial ring of star formation located to
the west of NGC 4038. 

(2) The seven most massive molecular gas complexes
are unusually large, with masses in the range of $3-6 \times 10^8$ 
M$_\odot$. These super-giant molecular complexes (SGMCs) 
are a factor of 5-10 times more massive
than the most massive giant molecular associations seen in the
grand-design spiral galaxy M51 (\markcite{rk90}Rand \& Kulkarni 1990).
Comparison of the CO map with the young star clusters
identified by \markcite{ws95}Whitmore \& Schweizer (1995) shows that
half of the extremely red clusters are located in the southern part of
the overlap region. This high
concentration of the red clusters towards the small fraction of the map
containing the brightest CO emission provides strong support for the
idea that the red color of the clusters is due to high extinction
(\markcite{ws95}Whitmore \& Schweizer 1995).

(3) Comparison of the CO data with the ISO 15 $\mu$m image shows that there
is an excellent correlation between the CO emission and the 15 $\mu$m
emission, particularly in the brightest regions. The three brightest
regions in the 15 $\mu$m image all have similar masses of molecular gas,
as well as similarly strong [NeIII]/[NeII] 
ratios, which indicate that all the regions are forming massive stars. 
However, the mid-infrared {\it continuum} spectrum is significantly
different from one region to another; in particular, the brightest
15 $\mu$m peak in the overlap region 
shows a strong and rising continuum spectrum, while the
other two bright peaks in this region show a relatively flat continuum
spectrum.

(4) We have presented two possible scenarios for explaining the presence
of strong, rising mid-infrared continuum emission in the SGMC4-5 region.
This continuum emission is produced by very small dust grains
heated to high temperatures by their proximity
to massive stars. These grains are expected to be removed from the 
neighborhood of the massive stars on relatively rapid timescales, most
likely by the effect of the stellar winds. One possibility is that the SGMC4-5
molecular complexes contain a larger number of young sites of star formation
than the other two, otherwise similar, large molecular complexes in
the overlap region, sites so young that the O stars have not yet
managed to blow away the dust. An alternative mechanism to produce the strong
continuum emission would be to replenish the dust grains around
older (few Myr) sites of star formation, perhaps via new material
being brought in by an ongoing collision of two massive molecular
complexes. Such collisions could
fuel the strong mid-infrared emission by continuously bringing fresh 
supplies of very small dust grains into the sphere of influence of
the massive stars that formed in the shocked boundary layer of the two
clouds. Although there is some support for the collisional scenario
in the kinematics and morphology of the CO emission in this region
of the galaxy, it is impossible with the current data to rule out
either scenario.

\acknowledgments

The research of CDW
is supported through grants from the Natural
Sciences and Engineering Research Council of Canada.  
The Owens Valley Millimeter Array is operated by the California
Institute of Technology and is supported by
NSF grant AST96-13717. VC would like to acknowledge the financial support
from a Marie Curie fellowship (TMR grant ERBFMBICT960967). We thank
the anonymous referee for comments which improved the presentation
of this paper.

\clearpage

\clearpage

\begin{table}
\dummytable\label{tbl-1}
\end{table}

\centerline{\psfig{file=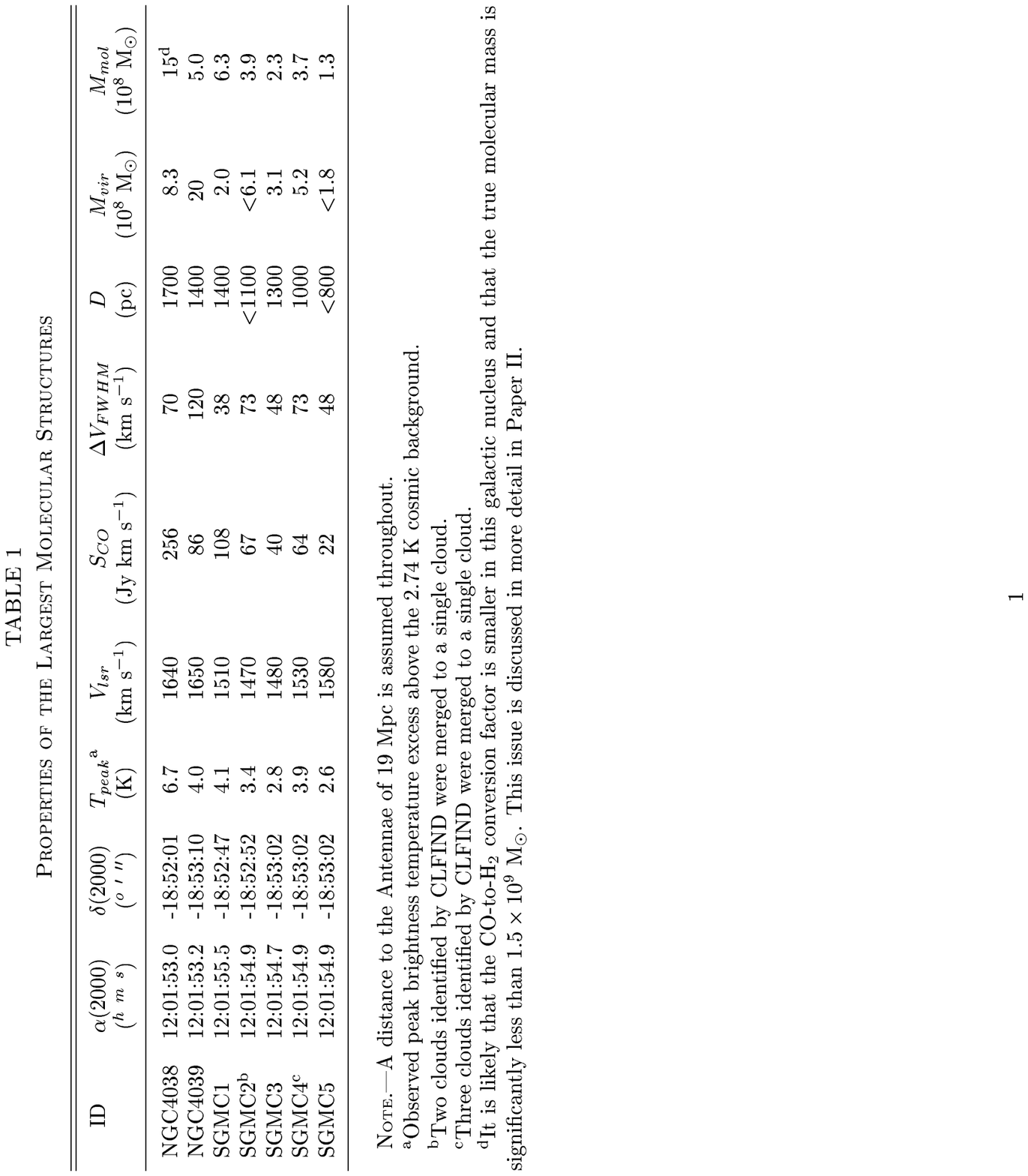,angle=90}}

\begin{deluxetable}{lccccccl}
\tablecaption{Mid-Infrared and CO Properties of Individual Complexes\label{tbl-2}}
\tablewidth{0pt}
\tablehead{
\colhead{ID} & \colhead{7 $\mu$m\tablenotemark{a}} & 
\colhead{15 $\mu$m\tablenotemark{a}}  & \colhead{$I_{CO}(peak)$\tablenotemark{b}} &
\colhead{15$\mu$m/7$\mu$m}  & \colhead{15$\mu$m/CO} &
\colhead{[NeIII]/[NeII]\tablenotemark{ac}} &
\colhead{Other IDs\tablenotemark{d}} 
} 
\startdata
NGC4038 & 28 & 38 & 91 & 1.4 & 0.42 & 0.1 & JK;7\nl 
NGC4039 & 15 & 24 & 32 & 1.6 & 0.75 & 0.5 & A;1 \nl 
SGMC1 & 27 & 43 & 38 & 1.6 & 1.1 & 1 & C;4;E;CD \nl 
SGMC2 & 29 & 39 & 38 & 1.3 & 1.0 & 1 & none;3;W;B \nl 
SGMC3 & 21 & 83 & 16 & 4.0 & 5.2 & 2 & B;2;S;A \nl
SGMC4,5\tablenotemark{e} & 28 & 150 & 51 & 5.4 & 2.9 & 4 &
none;2;S;A \nl 
\enddata
\tablenotetext{a}{Measured from the ISOCAM CVF spectrum in a
6$^{\prime\prime}$ pixel; units are mJy pixel$^{-1}$.}
\tablenotetext{b}{Peak CO intensity in a single $3.15\times 
4.91^{\prime\prime}$ beam; units are Jy km s$^{-1}$ beam$^{-1}$.}
\tablenotetext{c}{Note that there may be some small contamination
of the [NeII] line by the 12.7 $\mu$m UIB; this effect could cause 
the true line ratios to be higher by perhaps 10-20\% (see text).}
\tablenotetext{d}{Cross-identification with previous work at optical,
radio and infrared wavelengths. The letter and number designations refer
to \markcite{r70}Rubin et al. (1970), \markcite{h86}Hummel \& van der
Hulst (1986), \markcite{s90}Stanford et al. (1990), and \markcite{v96}Vigroux
et al. (1996), respectively.}
\tablenotetext{e}{SGMC4 and SGMC5 are combined in this table because
they are only separated in velocity, which the infrared data cannot
distinguish.}

\end{deluxetable}

\end{document}